\documentclass[prb,twocolumn,fleqn]{revtex4}
\usepackage{epsfig}
\usepackage{amsmath}
\usepackage{mathtools}
\usepackage{color}
\usepackage{graphics}
\usepackage{bm}
\usepackage{bbold}
\usepackage{dsfont}
\usepackage{changes}

\newcommand{\be}{\begin{equation}}
\newcommand{\ee}{\end{equation}}
\newcommand{\bea}{\begin{eqnarray}}
\newcommand{\eea}{\end{eqnarray}}

\newcommand{\subps}[3]{\ensuremath{{#1}_{\mbox{\tiny{#2}}}^{\mbox{\tiny{#3}}}}}
\newcommand{\ket}[1]{\ensuremath{\left|{#1}\right\rangle}}
\newcommand{\bra}[1]{\ensuremath{\left\langle{#1}\right|}}
\newcommand{\braket}[1]{\ensuremath{\left\langle{#1}\right\rangle}}
\newcommand{\blue}[1]{\textcolor{black}{#1}}
\newcommand{\unit}{\ensuremath{\mathds{1}}}
\newcommand{\summe}[2]{\sum\limits_{#1}^{#2}}
\newcommand{\void}[1]{}
\def\H{\mathcal{H}}
\def\Bad{\hat B^\dagger} \def\B{\hat B}
\def\aad{\hat a^\dagger} \def\a{\hat a}
\def\e{{\rm{e}}}
\def\i{{\rm{i}}}
\newcommand{\mynote}[1]{{\color{black}#1}}
\newcommand{\integral}[3]{\ensuremath{\int\limits_{#1}^{#2}{\rm d}#3\,}}
\newcommand{\preps}[3]{\ensuremath{\prescript{\mbox{\tiny{#3}}}{\mbox{\tiny{#2}}}{#1}}}
\renewcommand{\Re}[1]{\ensuremath{\operatorname{Re}\left(#1\right)}}

\begin{document}
\bibliographystyle{prsty}

\title{Apoptosis of moving, non-orthogonal basis functions
\\
in many-particle quantum dynamics}

\author{Michael Werther}
\affiliation{Max-Planck-Institut f\"ur Physik komplexer Systeme, N\"othnitzer Str. 38, D-01187 Dresden, Germany}
\affiliation{Institut f\"ur Theoretische Physik, Technische Universit\"at
Dresden, D-01062 Dresden, Germany}

\author{Frank Gro\ss mann}
\affiliation{Institut f\"ur Theoretische Physik,
Technische Universit\"at Dresden, D-01062 Dresden, Germany}

\begin{abstract}
Due to the exponential increase of the numerical effort with the number of degrees of freedom, moving basis functions have a 
long history in quantum dynamics. In addition, spawning of new basis functions is routinely applied.
Here we advocate the opposite process: the programmed removal of motional freedom of selected basis functions. 
This is a necessity for converged numerical results with respect to the 
size of a non-orthogonal basis, because
generically two or more states approach each other too closely early on, rendering unstable the matrix inversion, required 
to make the equations of motion explicit. Applications 
to the sub-Ohmic spin-boson model as well as to polaron dynamics in a 
Holstein molecular crystal model demonstrate the power of the proposed methodology.
\end{abstract}

\maketitle
\newpage
\section{Introduction}
\label{sec:intro}

The numerical effort for solving the time-independent and the time-dependent Schr\"odinger equation (TDSE) scales exponentially with the 
number of degrees of freedom. This is the reason why, up to the present date, \blue{one of} the largest molecular quantum systems whose dynamics can
be solved in an ab initio way in its full dimensionality, i.\ e., treating fully quantum mechanically all
the degrees of freedom by a suitable choice of fixed basis functions is the rather small, laser-driven hydrogen molecule 
H$_2$, consisting of just 4 particles \cite{PBM06}. Therefore, a lot of effort is devoted to the meticulous choice of
those fixed basis functions, with recent progress being made by using a small 
von Neumann basis of phase space Gaussians with periodic boundary conditions and biorthogonal exchange for the solution of the 
TISE for molecular problems \cite{ST12}.

In the TDSE case, much more flexible, however, are time-dependent basis functions, that move to and/or are created at positions where the support
of the wavefunction is. 
As reviewed below, they can be dealt with in a variational approach 
to the quantum dynamics as, e.g., in methods using coherent states, like Gaussian based multi-configuration time-dependent Hartree (G-MCTDH) methods \cite{ShBu08,KoFr13} 
as well as the Davydov-Ansatz \cite{DK73,Da73} and standard multi-configuration methods \cite{MMC90,BJWM00}. An in-depth review  of the variational multi-configurational Gaussian (vMCG) method 
with a discussion of numerical bottlenecks is given in \cite{Richings2015}. Furthermore, also moving position space grids have been considered, e.g., in the context of laser-driven dynamics of molecules \cite{LuBa01}.
An intriguing  possibility that has been explored for the basis function case is the creation of new such functions, for electronically non-adiabatic dynamics, 
whenever the wavepacket explores a new potential energy surface. If the forces for the classical dynamics of the parameters of the Gaussians
are calculated on the fly, this approach is called ab initio multiple spawning \cite{BQM00,MiCu18}. In general, all methods using Gaussian basis functions, due to their locality,
are well suited for on the fly dynamics as well as for treating finite temperature initial conditions. In the latter case, the $P$-function representation
of the canonical density operator may serve as a sampling density \cite{jcp19-02}.

In the present manuscript, we elaborate on an option that seems counterintuitive at first sight. This is the programmed removal
of a basis function's freedom, which we call apoptosis of basis functions, in contrast to the spawning alluded to above. Why would one want to do so? 
The reason is that the numerical stability of schemes that use non-orthogonal time-dependent basis functions to a large extend hinges on the possibility to render the equations to be 
discussed below explicit. To this end, some form of matrix inversion has to be applied \cite{Richings2015}. The matrix to be inverted becomes singular, however, in case two (or more) basis 
functions approach each other too closely, which generically happens close to convergence \cite{Haber12} and is referred to as linear dependency problem.

We will define a suitable measure for closeness and show that the removal of basis function freedom if that measure undershoots a certain threshold
leads to well-behaved numerics. Surprisingly, already a small number of basis functions is enough to obtain converged results for the full quantum dynamics of system \textit{and} 
environment in an open systems context. In the following, the open system is mimicked by discretizing the continuous spectral density of environmental oscillators using a suitable density of frequencies \cite{jcp19-02}. 
The method that we will employ to solve the TDSE of the composite system is the multi Davydov-Ansatz of type D2, developed in the Zhao group \cite{ZHZCZ15}.

The manuscript is structured as follows: First, in Sec.\ II, we introduce the methodological foundation for a generic many particle Hamiltonian and derive the equations of motion for the coefficients as well as the basis function parameters from a variational principle. In Sec.\ III, the Hamiltonian is specified to be of
system bath type, whereby the system of interest is treated using orthogonal basis functions. The treatment of the harmonic bath using coherent states in the present context then leads to the multi Davydov-Ansatz. After introducing our apoptosis strategy to circumvent the linear dependency problem close to convergence, this Ansatz serves as our workhorse for the solution of the dynamics  of two different model systems in Sec.\ IV: the spin-boson model, 
as well as the Holstein molecular-crystal model. 
Conclusions and an outlook are given in Sec.\ V. In the appendix,
remarks on the gauge freedom of the wavefunction Ansatz and
details of the regularization procedure, as well as
a convergence study for the spin-boson model can be found.

\section{Variational Coherent States Ansatz}\label{main_part}

We set the stage by first considering an $N$-particle Hilbert space and a dynamics being governed by the generic Hamiltonian 
\be
\label{eq:ham}
\hat{\mathcal H}=\sum_{j=1}^N\hat{H}_j+\sum_{i< j}\hat{W}_{ij},
\ee
with one-particle Hamiltonians $\hat H_j$ and  two-particle interactions $\hat W_{ij}$. 

An Ansatz for the solution of the TDSE is given in terms of 
multi-mode coherent states (CS) of multiplicity $M$ by
\bea
\ket{\subps{\Psi}{CS}{M}(t)}=\sum_{k=1}^M A_k(t)\ket{\bm{\alpha}_k(t)}, \label{eq:D_ansatz}
\eea
with time-dependent complex coefficients $A_k(t)$ and time-dependent $N$-dimensional complex displacements $\bm\alpha_k(t)$. 
$N$-mode CS are given by an $N$-fold tensor product
\bea
\ket{\bm{\alpha}_k} = \bigotimes_{j=1}^N\ket{\alpha_{kj}}
\eea
of normalized one-dimensional CS
\bea
\ket{\alpha_{kj}}=\exp\left[-\frac 12|\alpha_{kj}|^2\right]\exp\left[\alpha_{kj}\hat a_j^\dagger\right]\ket {0_j},  
\eea
where $\hat a_j^\dagger$ is the creation operator acting on the ground state
of a suitably chosen $j$-th harmonic oscillator and the CS form an over-complete and nonorthogonal basis set \cite{BBGK71}.
The generic Hamiltonian in (\ref{eq:ham}) is then to be expressed in terms of 
the creation and annihilation operators of the harmonic oscillator underlying 
the CS. In the cases to be considered below, the bath part of the Hamiltonian is
harmonic and this task is trivial.

The time-evolution of the coefficients and the displacements is governed by the Dirac-Frenkel variational principle \cite{Di30,Fren34}
\bea
\bra{\delta\subps{\Psi}{CS}{M}}\rm{i}\partial_t-\hat{\mathcal H}\ket{\subps{\Psi}{CS}{M}}=0,\label{eq:dirac_frenkel}
\eea
with $\hbar=1$ throughout the manuscript and where the variation reads
\bea
\bra{\delta\subps{\Psi}{CS}{M}}&=&\sum_{l=1}^M\bra{\bm{\alpha}_l}\left\{\delta A_l^\ast+A_l^\ast\sum_{j=1}^N\left[\left(-\frac 12\alpha_{lj}+\hat a_j\right)\delta\alpha_{lj}^\ast \right.\right.
\nonumber
\\
&-&\left.\left.\frac 12\alpha_{lj}^\ast\delta\alpha_{lj}\right]\right\}. \label{eq:D_variation}
\eea 
All appearing variations are mutually independent. Thus the equations of motion read
\bea
\bra{\bm{\alpha}_l}\text{i}\partial_t-\hat{\mathcal H}\ket{\subps{\Psi}{CS}{M}}&=&0, \label{eq:eom_1}
\\ 
A_l^\ast\bra{\bm{\alpha}_l}\hat a_j\left(\text{i}\partial_t-\hat{\mathcal H}\right)\ket{\subps{\Psi}{CS}{M}}&=&0, \label{eq:eom_2}
\eea
where the first equation was used to simplify the second one. 
These equations are similar to the vMCG ones \cite{Richings2015} but we use a novel solution strategy, detailed below.

By insertion of the explicit expression for the time-derivative of the Ansatz wave function
\bea
\partial_t\ket{\subps{\Psi}{CS}{M}}&=&\sum_{k=1}^M\Biggl\{\dot A_k+A_k\sum_{j=1}^N\left[-\frac 12\left(\alpha_{kj}\dot\alpha_{kj}^\ast+\dot\alpha_{kj}\alpha_{kj}^\ast\right)\right.
\nonumber
\\
&+&\left.\dot\alpha_{kj}\hat a_j^\dagger\right]\Biggr\}\ket{\bm{\alpha}_k}, \label{eq:D_timederivative}
\eea
equations (\ref{eq:eom_1},\ref{eq:eom_2}) can be solved in three steps. 
Firstly, to make progress, we introduce the combination of the 
time derivatives 
\be
\label{eq:x}
X_k:=\dot A_k +A_k\sum_{j=1}^N\left[-\frac 12\left(\alpha_{kj}\dot\alpha_{kj}^\ast+\dot\alpha_{kj}\alpha_{kj}^\ast\right)\right],
\ee
appearing in Eq.\ (\ref{eq:D_timederivative}),
as auxiliary variables, which is motivated by the gauge freedom inherent 
in the variational 
principle, as explained in more detail in Appendix \ref{app:gauge}.
Secondly, the linear system of equations for the $X_k$ as well as 
$\dot\alpha_{kj}$ is solved. To this end, we bring it into the form shown in Appendix \ref{app:lin}, without splitting real and imaginary parts.
The inversion problem is favorably tackled by using LU factorization with 
partial pivoting \cite{NUMREC}.
Thirdly, the obtained right hand sides of the equations for $\dot A_k$ 
and $\dot\alpha_{kj}$ are then used in the final step to integrate the highly nonlinear system of differential equations, favorably by using 
an adaptive Runge-Kutta method \cite{NUMREC}. 

Obviously, the second step above is problematic if the system matrix is 
(close to) singular, which is the case if either
\begin{itemize}
\item[(i)]
one of the coefficients $A_k\approx 0$ or if 
\item[(ii)] 
two CS approach each other too closely ($\bm\alpha_k\approx\bm\alpha_l$ for some $k\neq l$). 
\end{itemize}
This can most easily be seen by looking at the case $N=1$, for which system (\ref{eq:eom_1},\ref{eq:eom_2}) takes the form, see also \cite{ShBu08}
\bea
\text{i}\sum_{k=1}^M\braket{\alpha_l|\alpha_k}\Big[X_k&+&A_k\alpha_l^\ast\dot\alpha_k\Big]
\nonumber
\\
&=&\bra{\alpha_l}\hat{H}\ket{\subps{\Psi}{CS}{M}},\label{eq:eom_D_indep_N1_1}
\\
\text{i}A_l^\ast\sum_{k=1}^M\braket{\alpha_l|\alpha_k}\Big[\alpha_kX_k&+&A_k(1+\alpha_l^\ast\alpha_k)\dot\alpha_k\Big]
\nonumber
\\
&=&A_l^\ast\bra{\alpha_l}\hat a\hat{H}\ket{\subps{\Psi}{CS}{M}}.\label{eq:eom_D_indep_N1_2}
\eea
While for $A_k\approx 0$ one of the equations \eqref{eq:eom_D_indep_N1_2} turns into $0\approx 0$, for $\alpha_k\approx\alpha_l$ two of the equations \eqref{eq:eom_D_indep_N1_1} \textit{and} 
two of the equations \eqref{eq:eom_D_indep_N1_2} become approximately linearly dependent. 
We note in passing that canceling $A_l^\ast$ in the last equation is not appropriate for several reasons. 
Firstly, in the case $A_k=0$, the time-evolution of the corresponding CS $\ket{\alpha_k}$ can not be determined in terms of a first order differential equation \cite{UMan15}. 
Secondly, the inverse of the coefficient matrix corresponding to (\ref{eq:eom_D_indep_N1_1},\ref{eq:eom_D_indep_N1_2}) would not be unitary any more and norm conservation and stability would be lost. More
details can be found in Appendix \ref{app:lin}.

While the less severe first case (i) mentioned above may be treated by a regularization well-known from MCTDH \cite{Manthe1992}, and discussed in detail
in Appendix \ref{app:lin}, the second case (ii) is the more 
severe one known as the CS convergence issue \cite{Richings2015}. To put it pictorially: while the birth of a CS - accomplished by its equipment with an $\varepsilon$-sized 
coefficient - is well-behaved, it is not known how the death of a CS - desirable if two CS approach each other too closely, which generically happens close to convergence with respect to $M$ \cite{Haber12} - may be implemented. 
In order to circumvent this problem, various approaches such as re-expansion schemes \cite{KoFr13,Richings2015}, multiplication of the CS with orthogonal polynomials \cite{Hagedorn1981,Borrelli2016}, 
orthogonalizing momentum-symmetrized Gaussians \cite{PoSa04,LHT16} and projector splitting \cite{Bonfanti2018} have been applied. 

\section{The multi Davydov-Ansatz and Apoptosis}

Before we show how issue (ii) may be overcome, let us outline briefly how to 
apply the Ansatz \eqref{eq:D_ansatz} in a more general context.
If a ``system of interest'' of finite Hilbert space dimension $N_S$, e.\ g., a spin system 
is coupled to an environment of $N$ uncoupled harmonic oscillators 
\be
\hat H_j=\frac{\hat p_j^2}{2m_j}+\frac{1}{2}m_j\omega_j^2\hat x_j^2
,\qquad \hat{W}_{ij}=0,
\ee
the description of the environment by CS seems well justified. The equations \eqref{eq:dirac_frenkel} - \eqref{eq:eom_2} above may easily be extended to 
such a setting if an orthonormal basis 
$\left\{\ket{\phi_n}\,|\, n=1,\ldots,N_S\right\}$ of the system of interest's Hilbert space is chosen. The multi D2-Ansatz 
\bea
\ket{\subps{\Psi}{D2}{M}(t)}=\sum_{k=1}^M \left(\sum_{n=1}^{N_S}A_{nk}(t)\ket{\phi_n}\right)\ket{\bm{\alpha}_k(t)} \label{eq:D2_ansatz}
\eea  
then replaces \eqref{eq:D_ansatz}, and the equations of motion (\ref{eq:eom_1},\ref{eq:eom_2}) are replaced by
\bea
\bra{\phi_n}\bra{\bm{\alpha}_k}\text{i}\partial_t-\hat{\mathcal H}\ket{\subps{\Psi}{D2}{M}}&=&0,
\\
\sum_{n=1}^{N_S}A_{nk}^\ast\bra{\phi_n}\bra{\bm{\alpha}_k}\hat a_j\left(\text{i}\partial_t-\hat{\mathcal H}\right)\ket{\subps{\Psi}{D2}{M}}&=&0.
\eea
We stress that in the present D2 Ansatz, the coherent states do not carry the index $n$, in contrast to the so-called D1 Ansatz \cite{ Sun2010, Sun2015}.
In the MCTDH community the two approaches D2 and D1 are termed single and
multi set, respectively \cite{Richings2015}.
Furthermore, although the harmonic oscillators are not coupled directly to each other ($W_{ij}=0$), their combined wavefunction experiences non-Gaussian distortions due to the coupling to the spin system, 
requiring it to be represented by more than just a 
single multi-mode CS. As we will show below, the multiplicity $M$ of the D2 Ansatz, needed for convergence, is surprisingly low, however.

In order to tackle case (ii) mentioned above, we seek for a natural way to avoid the appearance of an ill-conditioned coefficient matrix
that causes $2$ of the equations \eqref{eq:eom_1} \textit{and} $2N$ of the equations \eqref{eq:eom_2} to become approximately linearly dependent. 
The system of equations being nonlinear, is expected to 
behave chaotically, but regularization of vanishing coefficients being successfully implemented, 
the system at the same time shows regular behavior. From this we conclude that it may be enough to remove  the linear dependencies in \eqref{eq:eom_2} only.
     
It is the linearity in the variations of \eqref{eq:D_variation} and the linearity in the displacements of \eqref{eq:D_timederivative} which is the key to implement this removal. 
To be more precise, assume that two CS $\ket{\bm{\alpha}_k}$ and $\ket{\bm{\alpha}_l}$ move from a certain time $t_0$ on \textit{connectedly}, i.e. without changing their relative position. 
Mathematically this means that the $N$ free parameters of one of them, say $\bm{\alpha}_l$, are replaced by the parameters of the other one as in the D1.5 Ansatz \cite{ps18}: 
\bea
\bm \alpha_l(t) = \bm{\alpha}_k(t)+\bm{C}, \label{eq:conn_disp}   
\eea
for $t\geq t_0$, where $\bm C=\bm{\alpha}_l(t_0)-\bm{\alpha}_k(t_0)$ is a constant. Consequently $\delta\alpha_{kj}=\delta\alpha_{lj}$ and $\dot\alpha_{kj}=\dot\alpha_{lj}$ for all $j$. 
At the level of the coefficient matrix, this amounts to deleting the $N$ rows/columns corresponding to the displacements $\alpha_{lj}$ and replacing the $N$ rows/columns corresponding to $\alpha_{kj}$ 
with the sum of both from time $t_0$ on.      
 
$\bm\alpha_l$ may from time $t_0$ on be regarded as dead, since its $N$ free parameters are removed. We name this programmed death for the ensemble's benefit apoptosis. Still the corresponding coefficient $A_l$ 
remains as a free parameter \cite{ps18}  which is highly advantageous, because, in contrast to a complete removal of the CS $\ket{\bm\alpha_l}$ \cite{Richings2015}, the norm of the Ansatz wave function is naturally conserved 
(no re-expansion is necessary) and no instabilities are introduced. 
Hence apoptosis is compatible with any adaptive integrator and can be done on the fly. 
Furthermore, keeping the coefficient comes at marginal computational cost since usually $M\ll N$. 

Depending on the precise problem to whose solution an Ansatz in terms of CS is 
used, the number $M$ of CS required to converge the problem may be large. 
In this case it may happen that multiple CS approach each other during 
propagation, and apoptosis of more than one CS could be required at a time step. 
Finding those CS which are close to each other can be implemented using a 
connected-component search in graphs \cite{Hopcroft1973}. 
Then, each connected component has to be replaced by one of its members only.

An important final detail concerns the position of those CS which are initially unpopulated, i.e., whose coefficients are initially zero. We may draw two conclusions from the above considerations. 
Firstly, those coefficients have to be subject to an initial noise due to case (i). This is in complete coincidence with the procedure used in MCTDH. Secondly, the precise position of those CS is in principle undetermined but should be governed by two restrictions: 
because of (ii) they should not come too close to any other CS, but on the other hand their distribution should be such that they represent unity, at least approximately. Both conditions can be fulfilled if the CS are centered around the initial condition on a multidimensional complex grid as given in \cite{BBGK71}.

\section{Applications}

In the following, we present applications to two problems that have proven
to be demanding test cases for several methods dealing with interacting 
many-body quantum systems, as there are path integral \cite{MSMT96,KA13} and 
multi-layer MCTDH methods \cite{WT10}, as well as renormalization group 
techniques \cite{BLTV05}, hierarchical equations of motion \cite{CZT15},
and tensor train propagation \cite{BG17}, to name but a few.

\subsection{Spin-Boson Dynamics}

First, we consider the symmetric spin-boson model at zero temperature \cite{Letal87},
\begin{align}
\hspace{-0.7cm}
\hat{\mathcal H}_{\text{\tiny SB}}=\frac{\Delta}{2}\hat\sigma_x -\frac{1}{2}\sigma_z\sum_{j=1}^N\lambda_j\left(\hat a_j^\dagger + \hat a_j\right) + \sum_{j=1}^{N}\omega_j\hat a_j^\dagger\hat a_j,
\end{align} 
where $\Delta$ is the tunneling amplitude and $\lambda_j$ is the coupling between the spin-1/2 system and the bath mode $\omega_j$. The relationship between the modes and their corresponding couplings 
is given by the spectral density (SD) of the bath oscillators, which we assume here to be of 
sub-Ohmic kind,  $J(\omega)=2\pi\alpha\omega_c^{1-s}\omega^s{\rm e}^{-\omega/\omega_c}$
with $s=0.25$, which is very demanding numerically. The Kondo-parameter $\alpha$ specifies the coupling strength, and $\omega_c$ is the high-frequency cutoff. 
Discretization of the SD is done via a density of frequencies $\rho_f\sim {\rm e}^{-\omega/\omega_c}$ \cite{jcp19-02}. 
In the following we take the two-state system initially to be in the state $\ket +$ and the bath to be equilibrated to the initial state of the two-state system\cite{KA13}:
\begin{align}
\ket{\Psi(0)}=\ket{+}\ket{\bm d},
\end{align} 
where $d_j=\frac{\lambda_j}{2\omega_j}$. Furthermore, we take $\omega_c$ to be the energy scale of the system and set $\Delta=-0.1\omega_c$. 
With these parameters, the model has been shown to support long lasting coherences \cite{KA13}.
\begin{figure}[ht]
	\includegraphics[scale=0.15,trim= 0cm 0cm 0cm 0cm,clip=True]{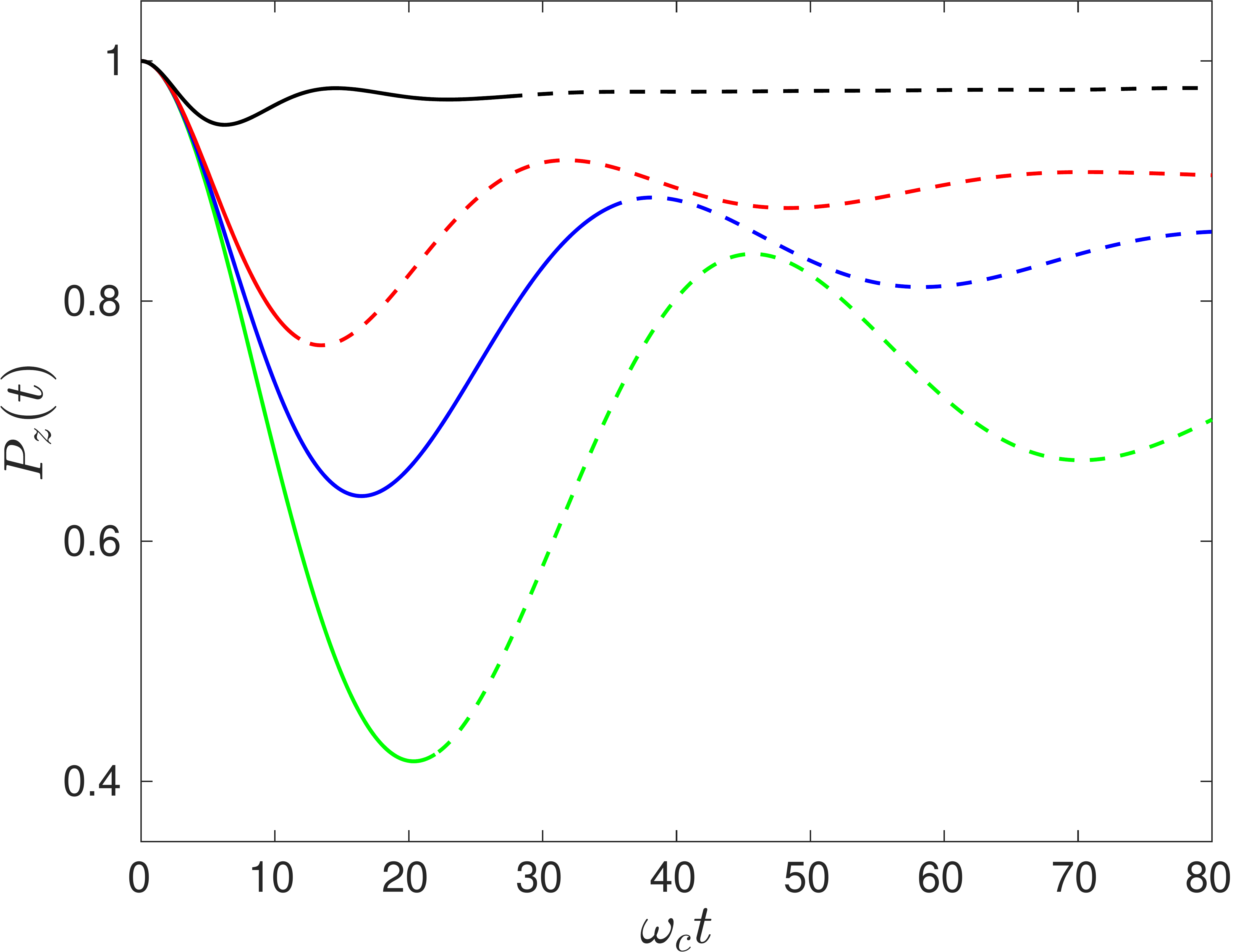}
	\caption{(color online) Dynamics of the population $P_z(t)=\langle\hat\sigma_z\rangle(t)$ of the spin-boson model with parameters given in the text for $N=150$ bath modes and multiplicity $M=10$. Apoptosis occurring for the first time  at times $\omega_c t$ (indicated by brokenness of line)
	for different coupling strengths:, $\alpha=0.03: \omega_c t=22.3$ (green line, lowermost curve),  $\alpha=0.04: \omega_c t=34.3$ (blue line, second-lowest curve), $\alpha=0.05: \omega_c t=12.8$ (red line, third-lowest curve), $\alpha=0.1: \omega_c t=28.6$ (black line, uppermost curve).}
\label{fig:comp_n}
\end{figure}

The result of the numerical implementation of the ideas laid out above is shown in Fig.\ \ref{fig:comp_n}. 
Apoptosis is applied if the distance $d(\ket{\bm\alpha_k},\ket{\bm\alpha_l})$ of two CS $\ket{\bm\alpha_k}$, $\ket{\bm\alpha_l}$ undershoots the threshold $\varepsilon=0.05$, which we found heuristically to be optimal for all tested systems. 
The distance $d$ is given by the $2$ product metric on $\mathbb{C}^N$, $d(\ket{\bm\alpha_k},\ket{\bm\alpha_l})=\sqrt{\sum\limits_{j=1}^{N}|\alpha_{kj}-\alpha_{lj}|^2}$. Without application of apoptosis, propagation for $\alpha=0.05$, with $M=10$, e.g., would be limited to the time-interval 
$\omega_c t\in[0,12.8]$, since at $\omega_c t\approx 12.8$ two CS come close, making the coefficient matrix nearly singular. With apoptosis implemented, propagation may be continued (dashed line) for times that are longer by an order of magnitude and beyond (not shown). 
It is remarkable that the number of CS coming close during propagation is not related to the multiplicity $M$ nor the coupling strength in an obvious way: propagation with increased $M$ may cope without apoptosis, or with more or fewer CS connected. 
Thus, in the presence of apoptosis convergence can be checked by increasing the multiplicity $M$ in a systematic way, either by increasing $M$ in a separate calculation starting again at time $t=0$ or by spawning new states on the fly.
A detailed convergence study is given in Appendix \ref{app:conv}.

In all the cases we have investigated, it turns out that the scaling of the numerical effort with respect to the number of degrees of freedom is extremely favorable.

\subsection{Polaron dynamics}

Secondly, for a molecular aggregate of $N$ molecules with periodic boundary conditions and one single electronic
two-level system per molecule, we have investigated the dynamics under the
Holstein molecular crystal model Hamiltonian, given by 
\cite{Sun2010}
\be
\H=\subps{\H}{ex}{} + \subps{\H}{ph}{} + \subps{\H}{int}{} \label{eq:Ham_exc_gen}
\ee
with diagonal coupling, where
\bea
\subps{\H}{ex}{}&=&-J\summe{n=-N/2+1}{N/2}\left[\Bad_n\B_{n+1}+\Bad_{n+1}\B_n\right], \\
\subps{\H}{ph}{}&=&\summe{n=-N/2+1}{N/2}\omega_n\aad_n\a_n, \\
\subps{\H}{int}{}&=&\summe{m,n}{}\lambda_n\omega_n\Bad_m\B_m\left(\a_n\e^{\i q_n m}+\aad_n\e^{-\i q_n m}\right).
\eea
Here, $\Bad_n$ and $\B_n$ are the exciton creation and annihilation operators of the $n$-th site, while $\aad_n$ and $\a_n$ are the creation and annihilation operators of a phonon of frequency $\omega_n$.
We consider a linear dispersion phonon band
\be
\omega(q)=\omega_0+W\left(\frac{2|q|}{\pi}-1\right).
\ee
By fixing even $N$ and taking the phonon momenta as
\be
q_n=\frac{2n\pi}{N},\qquad n=-\frac N2 +1,\ldots, \frac N2,
\ee
the corresponding frequencies are
\be
\omega_n=\omega_0+W\left(\frac{2|q_n|}{\pi}-1\right).
\ee
For this model we investigate two settings. In the first setting, the couplings $\lambda_n$
are constant,
\be
\lambda_n=\frac{g}{\sqrt N}\label{eq:const_coupl}
\ee
where $g$ is the diagonal coupling strength \cite{Chen2017}.
In the second setting, the couplings follow from the spectral density
\bea
J(\omega)&=&\frac{2S}{\pi W^2}\omega^2\sqrt{W^2-(\omega-\omega_0)^2}
\nonumber
\\
&\approx&\summe{n=1}{N}\lambda_n^2\omega_n^2\delta(\omega-\omega_n).\label{eq:SD_coupl}
\eea
Here, $S$ is the Huang-Rhys factor, $\omega_0=1$ is the central energy of the phonon band, and $W$ is the phonon energy bandwidth \cite{Sun2010}. 
In the following figures, we plot the diagonal elements of the exciton reduced density matrix 
\be
\rho_{nn}(t)=\bra{\Psi(t)}\Bad_n\B_n\ket{\Psi(t)}
\ee
as a function of $n$ and time. The exciton (two level system in the excited state) is initially at the middle position $n=0$, all the phonons are initially in their ground states. 

Our goal is to investigate events where two CS come close such that apoptosis is required. The number of these events is expected to be large in the regime where convergence with respect to the CS is reached. To be more precise: apoptosis is \texttt{always} 
required if the multiplicity $M$ is large enough, while 'large enough' depends on the setting. Indeed, in the first setting for $g=0.3$ where the couplings scale with $1/\sqrt{N}$ and are thus small, we find that already for $M>5$ the vast majority of propagations fails because of tiny integrator steps, 
if no apoptosis is applied. With apoptosis, the integrator always recovers, and each propagation successfully reaches the final time. Among those cases are many in which more than two CS are connected during propagation.  

Please note that the Hamiltonian \eqref{eq:Ham_exc_gen} as well as the initial state are symmetric with respect to site number. Thus, the CS which are initially unpopulated are also chosen such that they fulfill the symmetry. 
In this high-dimensional problem, no regularization of the $\rho$-matrix is required, and the coefficients of the CS which are initially unpopulated are set to $10^{-6}$.
 
\begin{figure}[ht]
	\includegraphics[scale=0.15,trim= 0cm 0cm 0cm 0cm,clip=True]{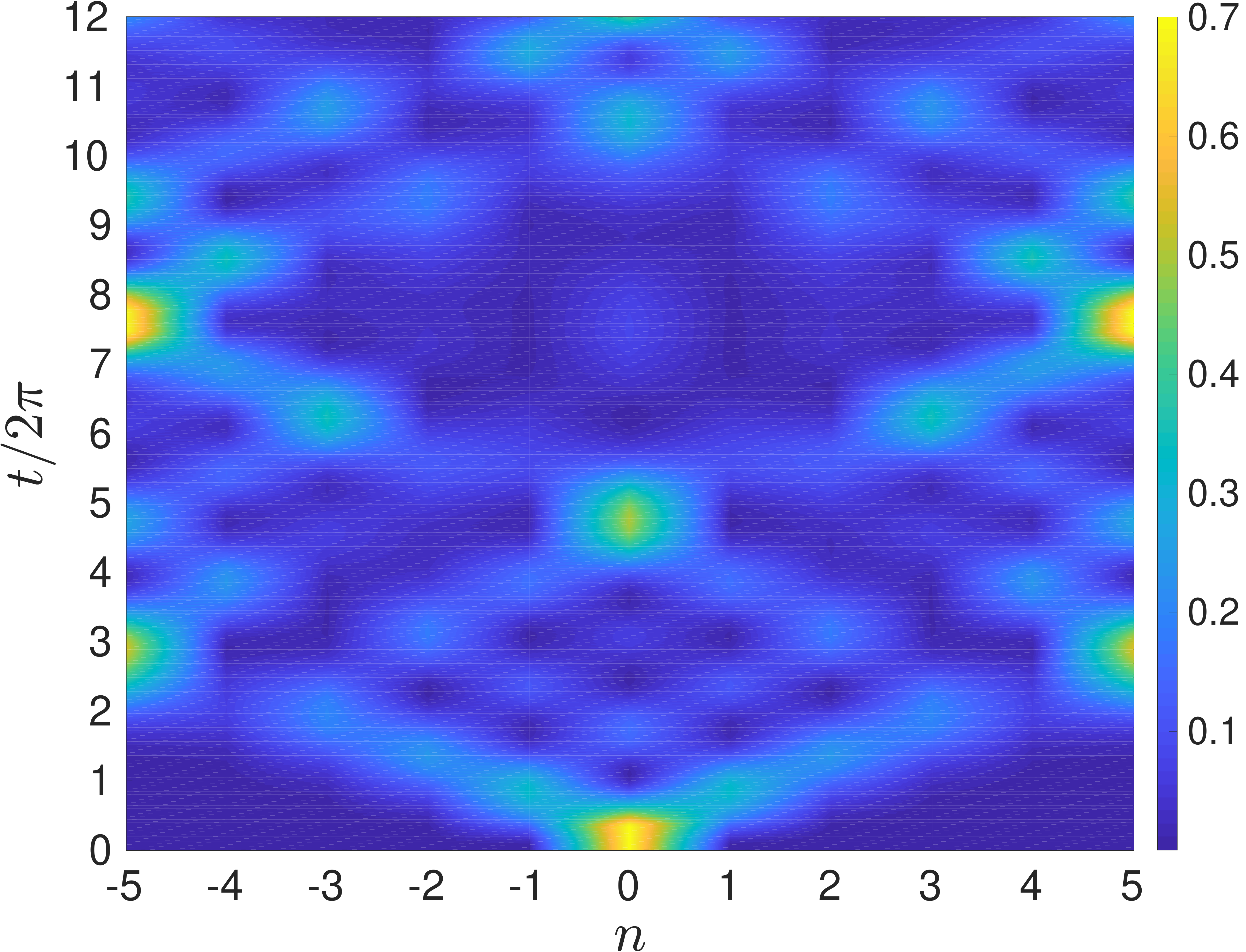}
	\caption{Reduced density matrix dynamics for the Holstein model with constant couplings \eqref{eq:const_coupl}. Parameters are  $g=0.3, J=0.2, W=0.5, N=10, M=9$. For the sake of better visibility of the dynamics at later times, $\rho_{nn}(t)$ has been restricted to $\rho_{nn}(t)\leq 0.7$ (i.e. if $\rho_{nn}>0.7$ then it is set to $0.7$).}
\label{fig:exc_1}
\end{figure}

In the first setting we extended to longer times the results of \cite{Chen2017}, where for constant couplings \eqref{eq:const_coupl} the parameters read $g=0.3, J=0.2, W=0.5, N=10$, corresponding to a model with 11 sites. 
In this case, already for small multiplicity, 
the results are fully converged. For instance the case $M=9$ is interesting,
because two CS come close right at the beginning of the propagation. Thus, without apoptosis, the propagation could not even start. With apoptosis applied, another event occurs at a later stage of propagation. The integrator recovers successfully from both events. The (converged) result for $M=9$ is shown in Fig. \ref{fig:exc_1}. 

\begin{figure}[ht]
\begin{centering}
\includegraphics[width=0.45\textwidth]{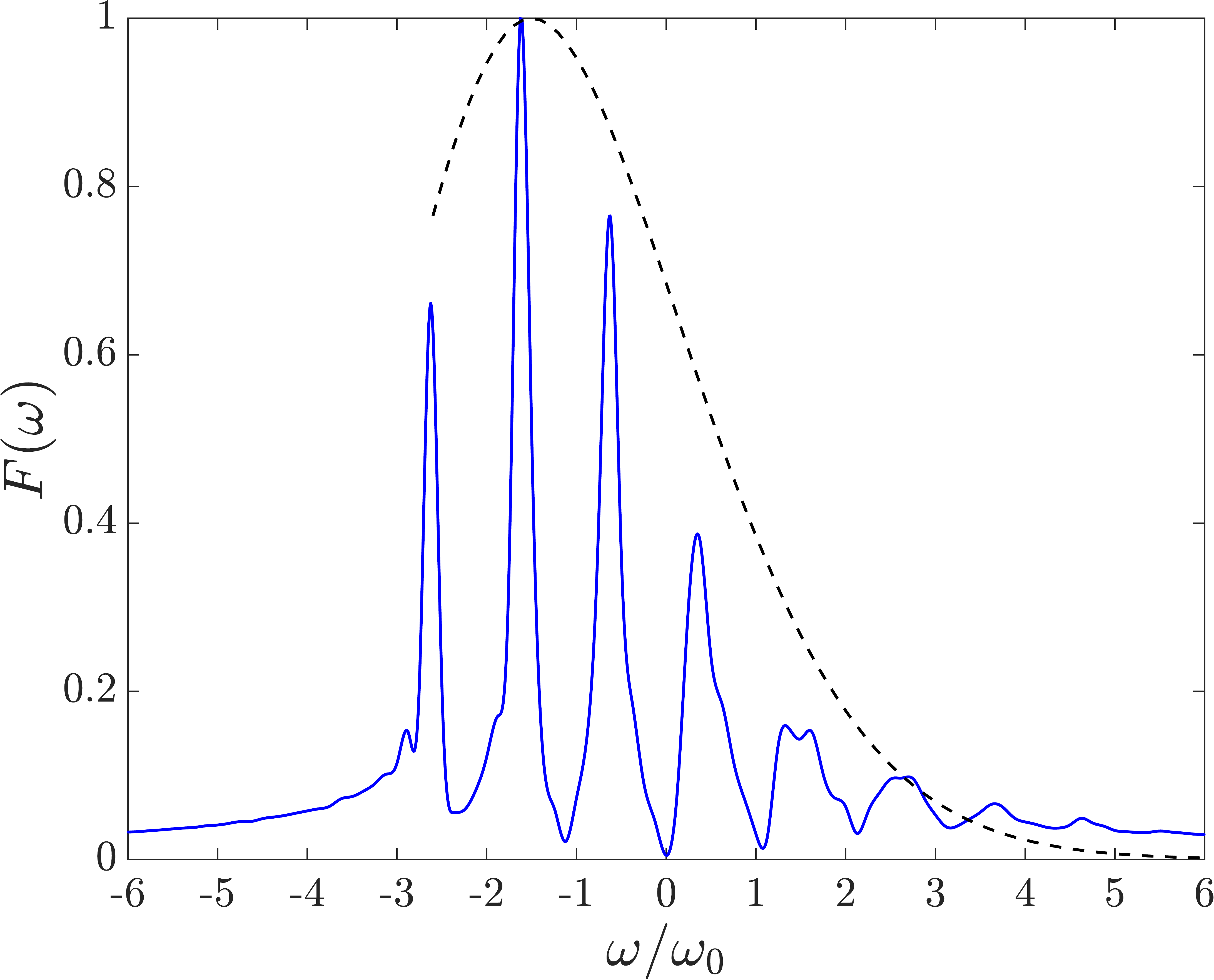}
\caption{The linear absorption spectrum as a function of $\omega$. The result obtained from the multi Davydov-Ansatz is plotted (blue solid) vs. the Poisson distribution \eqref{eq:Holstein_Poisson} (black dashed). The parameters read $N=16$, $J=0.1$, $W=0.1$, $g=0.4$.}
\label{fig_Holstein_abs_spec}
\end{centering}
\end{figure}

\mynote{Furthermore, again for the first setting, we compare the absorption spectrum from theoretical predictions with the Fourier transformed 
multi Davydov-Ansatz results. For a concise discussion of the
extraction of the spectrum from the dynamics, we refer to the appendix of \cite{ZHZCZ15}.  
The linear absorption spectrum for the parameter setting $N=16$, $J=0.1, W=0.1$ and $g=0.4$ is plotted in Fig. \ref{fig_Holstein_abs_spec}.
Huang-Rhys theory \cite{Huang1950} predicts the phonon side bands at zero temperature to follow a Poisson distribution,
\begin{align}
F(\omega)=\e^{-S}\summe{n=0}{\infty}\frac{S^n}{n!}\delta(\omega+S\omega_0-n\omega_0). \label{eq:Holstein_Poisson}
\end{align}
The leftmost sideband, $n=0$, is expected to be at $\omega=-S\omega_0$ where 
\begin{align}
S=\frac{1}{\omega_0}\summe{n=1}{N}\lambda_n^2\omega_n=\frac{Ng^2}{\omega_0}=2.56,
\end{align}
in nice coincidence with Fig. \ref{fig_Holstein_abs_spec}. Furthermore, the tallest peak is predicted to be at $n=S-1=1.56$, which again corroborates
our numerical result since the two peaks at $n=1$ and $n=2$ have similar height. Finally, by fitting a Poisson distribution with parameter $\lambda$ to the data, we find that the fit is optimal for $\lambda\approx S$ (see dashed black line in Fig. \ref{fig_Holstein_abs_spec}), which again confirms our results.}

\begin{figure}[ht]
	\includegraphics[scale=0.15,trim= 0cm 0cm 0cm 0cm,clip=True]{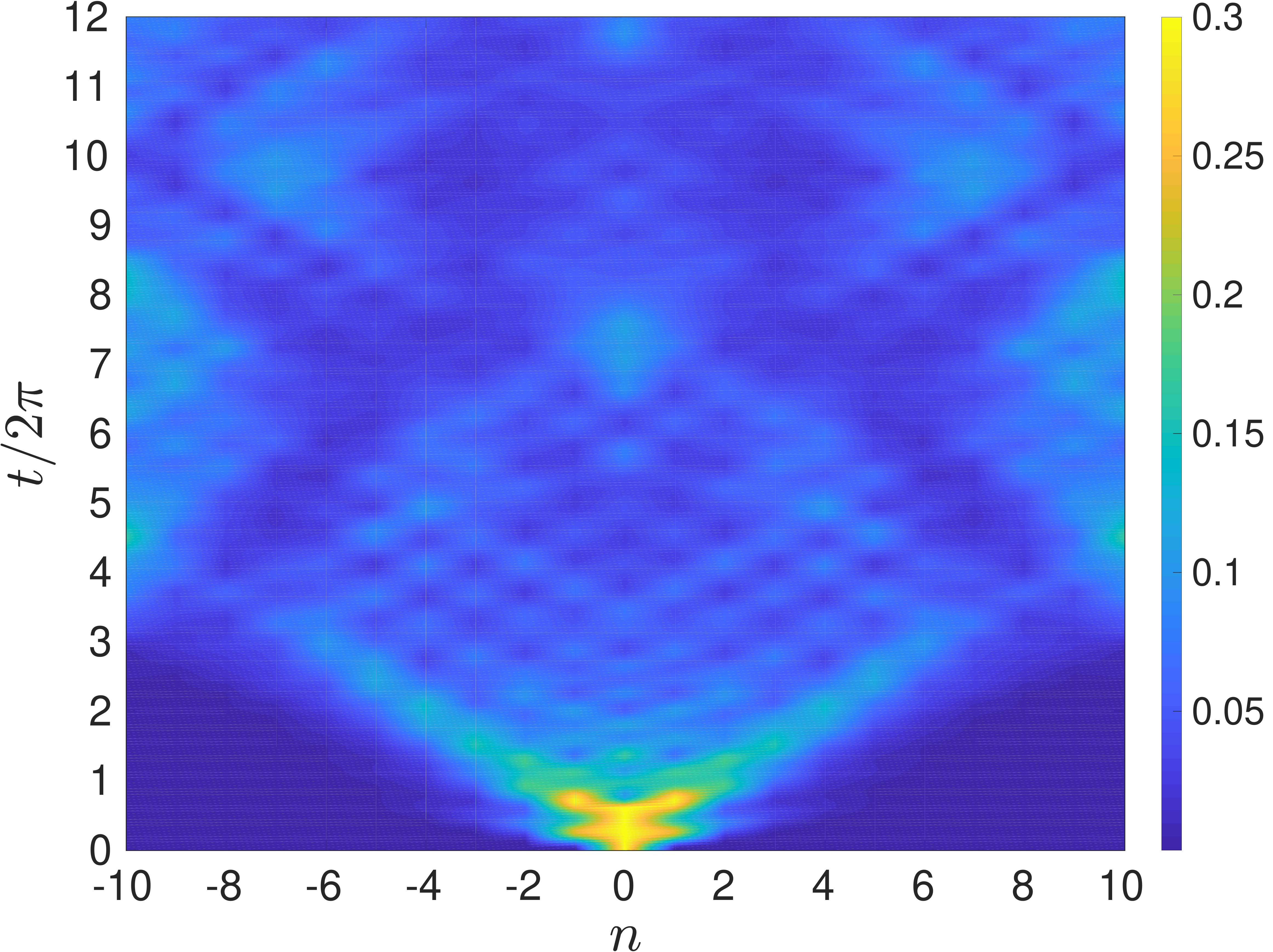}
	\caption{Reduced density matrix dynamics for the Holstein model of non-constant couplings given by \eqref{eq:SD_coupl}. Parameters are  $S=0.3, J=-0.5, W=0.8, N=20, M=30$. For the sake of better visibility of the dynamics at later times, $\rho_{nn}(t)$ has been restricted to $\rho_{nn}(t)\leq 0.2$ (i.e. if $\rho_{nn}>0.2$ then it is set to $0.2$).}
\label{fig:exc_2}
\end{figure}

In the second setting, we have extended to longer times and non-trivial multiplicity the results of \cite{Sun2010} where the couplings are given by \eqref{eq:SD_coupl}. 
There, we could not find events of CS coming close for parameters $N=30, S=0.5, W=0.8, J=-0.5$, for multiplicities up to $M=50$. This is expected since the coupling is rather strong (thus more CS would be needed for convergence). In a slightly modified setting ($S=0.3, W=0.8, J=-0.5, N=20$) 
we have found that the majority of propagations fails because of tiny integrator steps for $M>15$, if no apoptosis is applied. For instance for $M=20$, two CS come close at the very beginning of the propagation. The integrator recovers successfully with apoptosis, and the propagation ends with three CS connected. We find that the result is fully converged for $M=30$. Therein, apoptosis is needed since two CS come close at $\frac{t}{2\pi}\approx 5.78$. Again, the integrator recovers successfully from the apoptosis-event. The result is shown in Fig. \ref{fig:exc_2}. 

\section{Conclusions and Outlook}

We have shown that the temporal stability of the numerics for many-particle quantum dynamical simulations using time-dependent coherent states can
be enhanced dramatically by using apoptosis, i.\ e., programmed removal of basis function freedom. For 150 oscillators in a spin-boson dynamics,
that, by using orthogonal basis functions, only multi-layer MCTDH methods could cope with so far \cite{WT10}, a small double digit multiplicity of moving Gaussians was enough to achieve 
converged results for sub-Ohmic spectral densities and several oscillation periods of the spin system. 
Also the exciton dynamics in a Holstein molecular crystal model can be
converged using small multiplicities, especially in the case of constant
coupling.

The key to the long-time stability of our approach, apart from apoptosis,
is the use of normalized coherent states, the introduction of the auxiliary
variables $X_k$ for the solution of the linear algebra inversion problem, 
and regularization of the so-called $\rho$-matrix, as detailed in 
Appendix B.
Technically, the compatibility of apoptosis with the integrator would  
allow to reverse the procedure by connecting two CS at some \textit{large} distance and freeing them again at a later stage 
of the propagation, something we would like to investigate in the future. 
In addition, the presented approach is not restricted to 
problems with a finite-dimensional Hilbert space of the system of interest.
Also implementations for degrees of freedom with a continuous 
variable will benefit from the proposed numerical scheme, as we will show in
a future publication.

Finally, also finite temperatures of the bosonic heat bath can be accounted for 
by additional initial condition sampling using a $P$-function representation of
the canonical density operator \cite{GaZo}.

{\bf Acknowledgments:}
The authors are grateful for enlightening discussions with Robert Binder,
Matteo Bonfanti, Irene Burghardt, Richard Hartmann, Uwe Manthe, 
Walter Strunz, and Yang Zhao.
FG would like to thank the Deutsche Forschungsgemeinschaft for financial support under grant GR 1210/8-1.

\begin{appendix}

\section{Gauge Freedom}\label{app:gauge}

It is central to our approach to solve the variational equations of motion 
that there is a gauge freedom in the Ansatz (\ref{eq:D_ansatz}), which is invariant with respect to (time-dependent) linear transformations of the CS basis. Let ${\bf Q}$ be a nonsingular transformation matrix, then the wave function remains unchanged if 
$A_k$ and $\ket{{\bm \alpha}_k}$ are replaced by
\bea
A_k\to\tilde A_k&=&\sum_{l=1}^{M}A_l\left({\bf Q}^{-1}\right)_{lk},
\\
\ket{{\bm \alpha}_k}\to\ket{\tilde {\bm \alpha}_k}
&=&\sum_{l=1}^{M}Q_{kl}\ket{{\bm \alpha}_l}.
\eea
This is analogous to the multi-configurational time-dependent Hartree approach 
\cite{MMC90,Manthe1992}, where the gauge freedom is used to significantly simplify the equations of motion. \\
For diagonal transformations ${\bf Q}$, the procedure effectively amounts to multiplication of each CS with a possibly time-dependent non-zero C-number. The transformation
\be
Q_{kl}:=
\left\{
\begin{array}{ll}
\exp\left[\frac{1}{2}\sum_{j=1}^N|\alpha_{kj}|^2\right],& k=l 
\\ 
0,& k\neq l
\end{array}
\right .
\ee
results in the same wave function but with unnormalized (Bargmann) CS. Then, solving the linear system for these and transforming back is equivalent with transforming the time derivatives of the coefficients forth and back according to
\be
X_k=Q_{kk}\partial_t\left[A_k\left({\bf Q}^{-1}\right)_{kk}\right],
\ee
which is equivalent to Eq.\ (\ref{eq:x}).
The disadvantage of using the equations with unnormalized CS from the start is that the coefficients may become large, which is not the case if normalized CS are employed \cite{dissMW}. 

The introduction of the $X_k$ variables allows to both use normalized CS and have the linear algebraic system of equations in standard form (see Appendix \ref{app:lin}). This is the preferred way of dealing with the appearance of 
$\dot{\alpha}_{k}$ as well as $\dot{\alpha}_{k}^\ast$ in Eq.\ (\ref{eq:D_timederivative}).

\section{Regularization Details}\label{app:lin}

In the following, we detail how to disentangle instabilities arising due to closeness of coherent states from those arising due to almost vanishing coefficients. \mynote{The general system of equations of motion emerging from (\ref{eq:eom_1},\ref{eq:eom_2}) reads 
\bea
\i\sum_{k=1}^M\Big\{X_k&+&A_k\summe{n=1}{N}\dot\alpha_{kn}\alpha_{ln}^\ast\Big\}\braket{\bm\alpha_l|\bm{\alpha}_k}\notag \\ &=&\bra{\bm{\alpha}_l}\hat{\mathcal H}\ket{\subps{\Psi}{CS}{M}},\label{eq:eom_CS_general_wavefunction_X_1}
\\ 
\i A_l^\ast\sum_{k=1}^M\Big\{\alpha_{kj}\big(X_k&+&A_k\summe{n=1}{N}\dot\alpha_{kn}\alpha_{ln}^\ast\big)+A_k\dot\alpha_{kj}\Big\}\braket{\bm\alpha_l|\bm{\alpha}_k} \notag \\ &=&A_l^\ast\bra{\bm{\alpha}_l}\a_j\hat{\mathcal H}\ket{\subps{\Psi}{CS}{M}}, 
\label{eq:eom_CS_general_wavefunction_X_2}
\eea
with the notations used in the main text.} We set 
\bea
{\bf x}&:=&\left(X_1,\ldots,X_M\right), 
\\
{\bf y}&:=&\left(\dot\alpha_{11},\ldots,\dot\alpha_{MN}\right), 
\\
{\bf A}&:=&\left(A_1,\ldots,A_M\right).
\eea
Then the linear system (\ref{eq:eom_CS_general_wavefunction_X_1},\ref{eq:eom_CS_general_wavefunction_X_2}) then takes the 
standard form
\be
\label{eq:stand}
{\rm i}\begin{pmatrix} \mathbf S & \mathbf B \\ \mathbf B^\dagger & \mathbf D \end{pmatrix}\begin{pmatrix} {\bf x}^{\rm T}\\ {\bf y}^{\rm T} \end{pmatrix}
=
\begin{pmatrix} {\bf r} \\ {\bf s} \end{pmatrix},
\ee
where
\be
S_{lk}=\langle\bm\alpha_l|\bm\alpha_k\rangle,
\ee
are elements of the \mynote{Hermitian} $M$$\times$$M$ overlap matrix and
\be
\mathbf B= [\mathbf F^\ast \otimes {\bf A}]
\circ[\mathbf S\otimes 1_{1\times N}] ,
\ee
is an $M$$\times$$NM$ matrix, whereas 
\bea
\mathbf D &=& \left(\left[1_{1\times N}\otimes\mathbf F^{\rm T}\otimes 
1_{M\times 1}\right]\circ\left[1_{N\times 1}\otimes\mathbf F^\ast\otimes
1_{1\times M}\right]\right)
\nonumber\\
&&\circ\left[1_{N\times N}\otimes\left(\bm\rho\circ\mathbf S\right)\right]+ 
\unit_{N}\otimes\left(\bm\rho\circ\mathbf S\right)
\eea
is a \mynote{Hermitian} $NM$$\times$$NM$ matrix \footnote{Note that 
$\left[1_{1\times N}\otimes\mathbf F^{\rm T}\otimes 
1_{M\times 1}\right]^\dagger=\left[1_{N\times 1}\otimes\mathbf F^\ast\otimes
1_{1\times M})\right]$, simplifying the numerical calculation of $\mathbf D$.} for whose derivation we had to employ the  anticommutation relation $[\hat{a}_j,\hat{a}_j^\dagger]=\hat 1$.
In addition, we have used the $M$$\times$$M$ \mynote{single-particle density} matrix 
$\bm\rho = \left({\bf A}^\dagger\otimes {\bf A}\right)$ known from 
MCTDH \cite{Manthe1992},
and the $M$$\times$$N$ matrix of displacements $\mathbf F=(\alpha_{kj})$. 
Furthermore, $\otimes$ denotes the tensor-product 
and $\circ$ the Hadamard-product (element-wise multiplication), $1$ are 
matrices of ones and $\unit$ is the identity matrix for the indexed dimensionality, whereas a dagger
denotes Hermitian conjugation. 

\mynote{The right hand side of Eq.\ (\ref{eq:stand}) 
is given by
\bea
{\bf r}&=&[{\bf H}\circ{\bf S}]{\bf A}^{\rm T}
\\
{\bf s}&=&\mbox{vec}\left[\Big(\bm\rho\circ\mathbf S\circ\mathbf H\Big)\mathbf F+\left(\Big(1_{1\times 1\times N}\otimes\big(\bm\rho\circ\mathbf S\big)\Big)\circ\tilde{\mathbf H}\right)_2\right] \notag \\
& &
\eea
where we have assumed a normally ordered Hamiltonian and ${\bf H}$ 
is the matrix with elements $H_{\rm ord}(\bm\alpha_l^\ast,\bm\alpha_k)$, 
whereas the tensor $\tilde{\bf H}$ has the  
elements $\tilde{H}_{lkn}=\frac{\partial H_{\rm ord}(\bm\alpha_l^\ast,\bm\alpha_k)}{\partial\alpha_{ln}^\ast}$ (see, e.g.\ \cite{ShBu08}). Furthermore, $\mbox{vec}\big[\cdot\big]$ denotes the vectorization\footnote{The vectorization $\mbox{vec}\big[\mathbf M\big]$ turns a $m\times n$ matrix $\mathbf M$ column-wise into a $mn\times 1$ column vector according to \\ $\mbox{vec}\big[\mathbf M\big]=\left(M_{11},\ldots,M_{m1},\ldots,M_{1n},\ldots,M_{mn}\right)^T$.} of a matrix and $\big(\cdot\big)_2$ denotes summation over the second index.} The generalization of this exposition to the Davydov case
can be found in \cite{dissMW}. \mynote{We note in passing that while $\tilde{\mathbf H}$ in general has tensorial character, it often simplifies tremendously, as e.g. for the case of a set of mutually uncoupled oscillators in an open system context, where $\hat H=\sum\limits_n\omega_n\aad_n\a_n$ and thus $\tilde{H}_{lkn}=\omega_n\alpha_{kn}$.}

Clearly, in general also the block $\mathbf B$ is decisive for regularity of the full matrix, but our implementations show that no further instabilities arise once $\mathbf S$ and $\mathbf D$ are sufficiently regular. The closeness of coherent states endangers the regularity of $\mathbf S$ 
(and also of $\mathbf F$), while vanishing coefficients endanger the regularity of $\mathbf D$. While we have outlined in the main article how to solve the first issue by apoptosis, we will detail now how to regularize $\mathbf D$.  

In a first attempt we have tried to regularize $\mathbf D$ by replacing it with $\mathbf D+\delta\exp\left[-\bm\rho/\delta\right]$ for $\delta\ll 1$. This lead to further instabilities, most likely because this influences also the displacements. 
In view of the special structure of $\mathbf D$ and keeping in mind that apoptosis ensures regularity of $\mathbf F$, it turns out that it is much more expedient to regularize $\bm\rho$ only (this being the main reason behind not cancelling $A_l^\ast$ in Eq.\ (\ref{eq:eom_D_indep_N1_2})). This is done by replacing 
$\bm\rho$ by either 
$\bm\rho+\varepsilon_\rho\exp\left[-\bm\rho/\varepsilon_\rho\right]$ 
(see, e.g., \cite{Manthe1992}) or even by $\bm\rho+\varepsilon_\rho\unit_M$ for $\varepsilon_\rho\ll 1$. This indeed does not effect the displacements, but effects the coefficients (belonging to nearly unpopulated coherent states) only. 

Finally, our implementations show that a strong regularization of $\bm\rho$ is required in low-dimensional problems, especially if many coherent states are propagated. On the contrary, even if many coherent states are propagated, (almost) no regularization of $\bm\rho$ is required in high-dimensional problems.

\section{Convergence study for the spin-boson case}\label{app:conv}

For an intermediate coupling strength of $\alpha=0.04$, we study in detail the convergence of the numerical results for the spin-boson model discussed in the main text. To this end, we define an error measure
\be
\Delta_A=\frac{1}{N_t}\sum_{i=1}^{N_{t}}|P_{A_{\rm m}}(t_i)-P_A(t_i)|
\ee
where $P$ is the population displayed in Fig.\ 1  and with $N_t=500$ points 
in time that are spaced equidistantly. Although our numerical integrator 
uses adaptive time steps, the output is given at equidistantly spaced points.

The quantity $A$ indexing $P$, with respect to which convergence is checked, can be either (i) $N$,
which is the number of bath oscillators of the spin-boson model with discretized spectral density, or (ii) $M$,
which is the multiplicity of the D2 Ansatz. With $A_{\rm m}$, we denote the maximum value of 
the parameter that we have chosen (for which convergence of the numerical results in the plot shown 
in the article to within line thickness is reached).

\begin{figure}[ht]
	\includegraphics[width=0.45\textwidth]{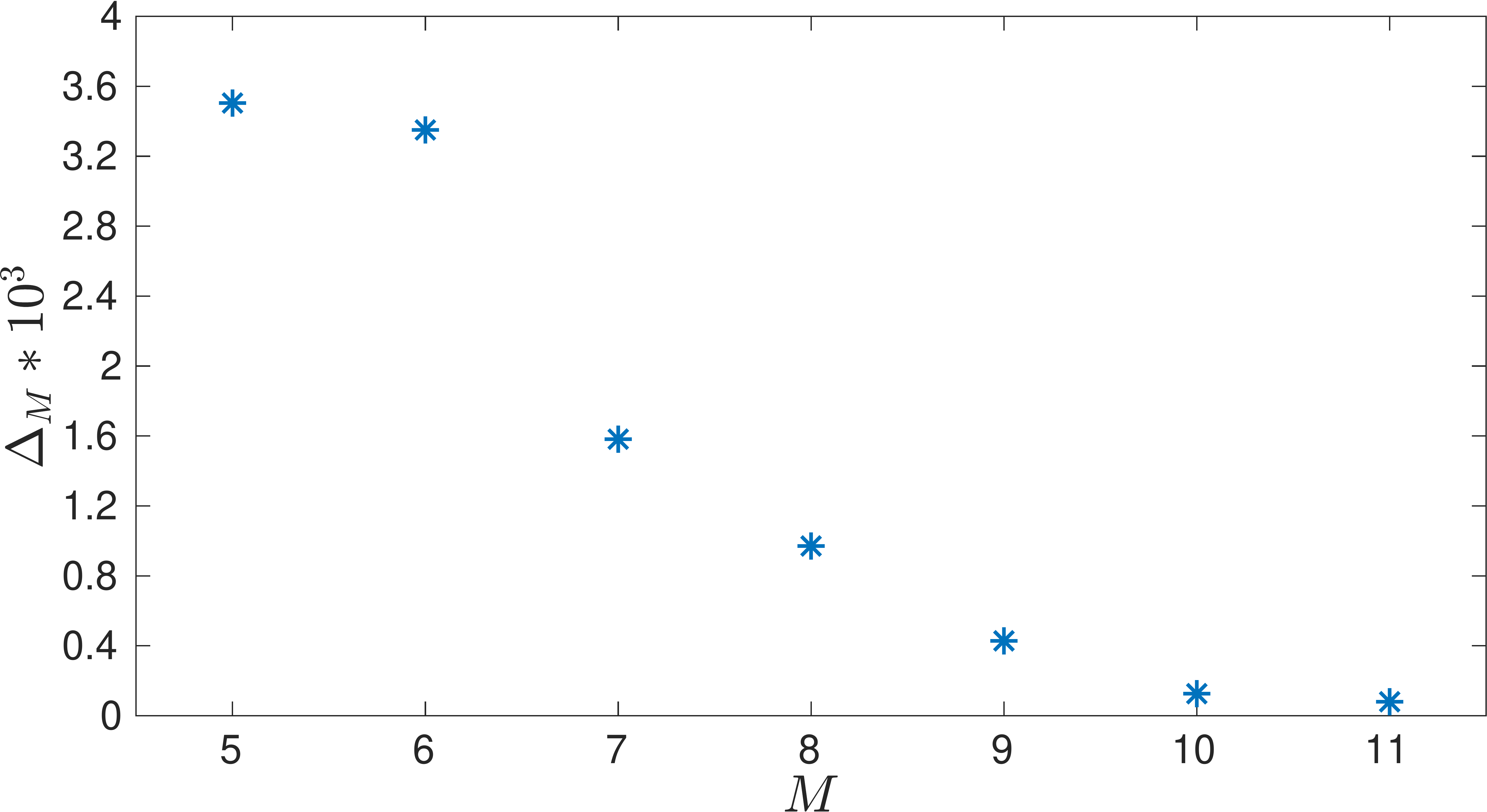}
	\caption{Convergence with respect to multiplicity $M$ of the multi Davydov-Ansatz for $N=150$
	oscillators chosen according to the discretization method mentioned in the original article
	and for coupling strength $\alpha=0.04$. }
\label{fig:conv_M}
\end{figure}

In Fig.\ \ref{fig:conv_M} for a number of $N=150$ bath modes, the convergence with respect to the multiplicity is
checked. Using $M_{\rm m}=12$, it turns out that $M=10$ leads to the converged results shown in our paper that coincide exactly with
the ones from \cite{KA13}.

\begin{figure}[ht]
	\includegraphics[width=0.45\textwidth]{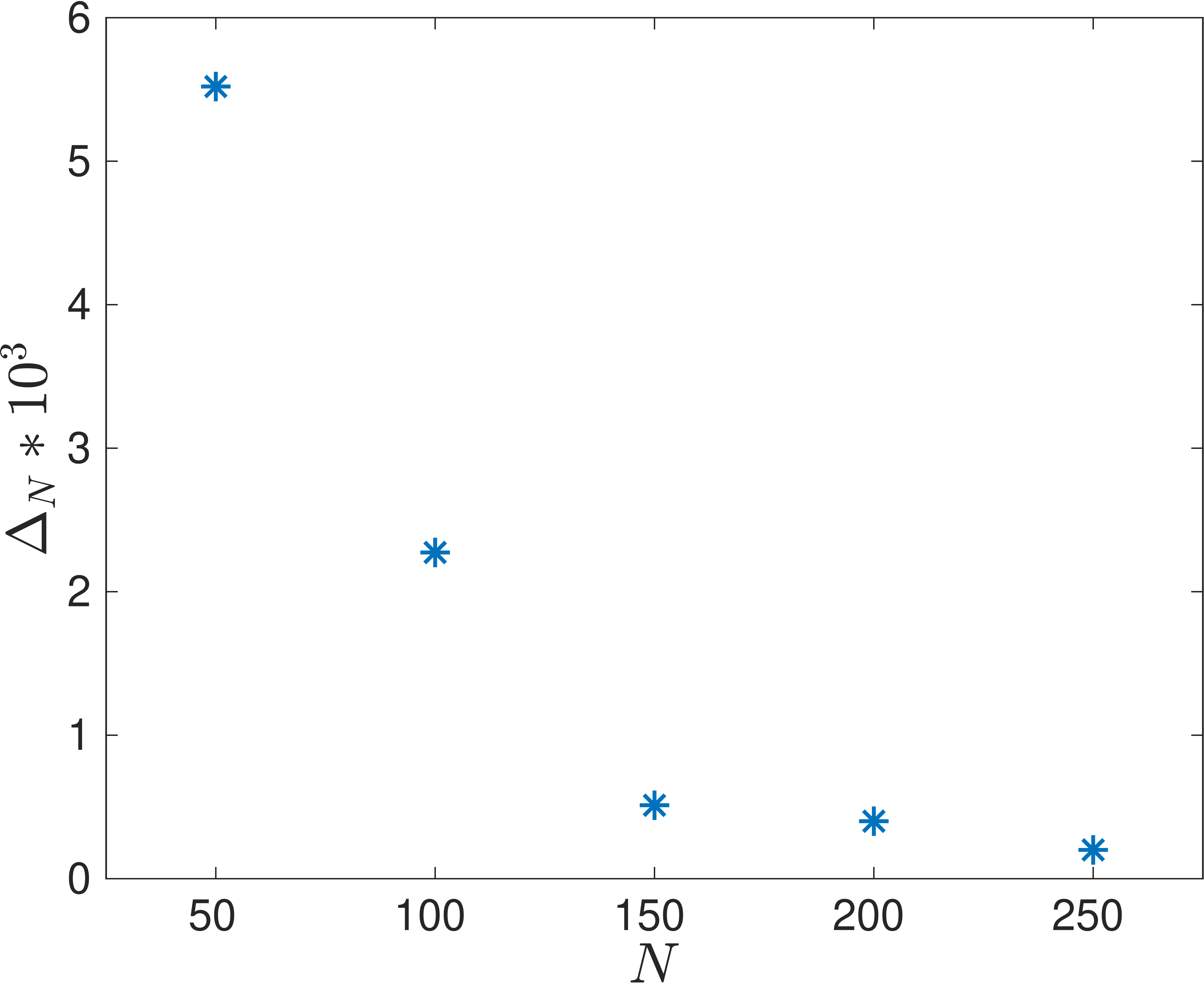}
	\caption{Convergence with respect to $N$ for a multiplicity $M=10$ and for coupling strength $\alpha=0.04$. }
\label{fig:conv_N}
\end{figure}

In Fig.\ \ref{fig:conv_N} for a multiplicity of $M=10$, the convergence with respect  to the number of bath modes $N$ is
checked. Using $N_{\rm m}=300$, it turns out that $N=150$ leads to the converged results shown in our paper that coincide with
the ones from \cite{KA13}.

Although the convergence check is possible self-consistently, it helped a lot to have 
the converged results of \cite{KA13} at our disposal. In this respect it is intriguing that 
the same values of parameters $M$ and $N$ that lead to convergence for coupling strength $\alpha=0.04$
are also suitable for the other coupling strengths considered. In passing, we note that the initial choice of the 
centers of the unpopulated Gaussians plays a minor role as long as they are distributed close enough
around the initial condition.

Finally, it is worth mentioning that the error would increase if the time series was extended to longer times,
which could, however, be cured by starting out with a higher multiplicity or by spawning new CS at a later stage of propagation.

\end{appendix}

\section*{References}


\end{document}